\documentclass[]{cupconf}
\usepackage{graphicx}

\title[Solar Chemical Peculiarities?]
      {Solar Chemical Peculiarities?}
      
\author[C. Allende Prieto]{\\ Carlos Allende Prieto$^1$} 
\affiliation{$^1$ McDonald Observatory and Department of Astronomy, 
University of Texas, Austin,  USA}   

\begin{document}
\maketitle

\begin{abstract}
Several investigations of FGK stars in the solar neighborhood have suggested  
that thin-disk stars with an iron abundance similar to the Sun appear to 
show higher abundances of other elements, such as silicon, titanium, or nickel. 
Offsets could arise if the samples contain stars
with ages, mean galactocentric distances, or kinematics, 
that differ on average from the solar values. They could also arise due 
to systematic errors in the abundance determinations, if the samples 
contain stars that are different from the Sun regarding their atmospheric 
parameters. We re-examine this issue by studying a sample 
of 80 nearby stars with solar-like colors and luminosities.
Among these solar {\it analogs}, the objects  with solar iron abundances 
exhibit solar abundances of carbon,  silicon, calcium, titanium 
and nickel.
\end{abstract}

\firstsection
\section{Introduction}

Under the assumption that low-mass dwarf stars with convective envelopes 
have a surface chemical composition which simply reflects that of their
natal clouds, one can use such stars to trace the chemical evolution of
the Galaxy. The Sun is then a convenient reference, but are there nearby
stars with solar composition? or in other words, is the solar abundance
pattern the norm in the local thin disk? 

Inspection of some of the most recent synoptic studies of nearby stars
suggests that the Sun's metallicity is slightly off from average  
(metallicity is here equated to the iron abundance 
[Fe/H]
\footnote{[Fe/H]$\displaystyle = \log_{10} \frac{\rm N(Fe)}{\rm N(H)} +12$, 
where N represents number density}). 
Nordstr\"om et al. (2004) obtained metallicities from Str\"omgrem photometry
for nearly 14,000 F- and G-type stars within 70 pc, finding that their
distribution could be approximated by a Gaussian with a mean of 
[Fe/H]$=-0.14$ and a $\sigma$ of 0.19 dex.
Allende Prieto et al. (2004) studied spectroscopically the stars more 
luminous than $M_V=6.5$ ($M>0.76 M_{\odot}$) within 14.5 pc from the Sun 
and concluded that their metallicity distribution is 
centered at [Fe/H]$=-0.11$ and has a $\sigma$ of 0.18 dex. 
Luck \& Heiter (2005) derived spectroscopic metallicities 
for a sample of 114 FGK stars within 15 pc (similar to that analyzed by  
Allende Prieto et al. 2004), finding a metallicity distribution with a
consistent width ($\sigma = 0.16$ dex), but centered at a value slightly 
closer to solar ($-0.07$ for the complete sample, and $-0.04$ 
when thick-disk stars are excluded). 
Haywood (2002) has argued that sample
selection based on spectral type discriminates against high-metallicity
stars, proposing a metallicity distribution 
for the solar neighborhood (based on photometric indices) 
that is centered at the solar value.

Inevitably, one must ask whether there is any reason to expect the
local metallicity distribution to be centered at the solar value.
Chemical differences among the Sun and its neighbors may be
reasonable if the age or the Galactic orbit of the Sun are
somewhat off from the average for nearby stars. 
The age distribution or, equivalently, the star formation history of
the solar neighborhood is an unsolved problem, judging
from the discrepant results obtained from analyses of the Hipparcos
H-R diagram (Bertelli \& Nasi 2001, Vergely et al. 2002) and 
studies of stellar activity (e.g. Rocha-Pinto et al. 2000).

What about abundance ratios? Should we expect the ratios 
such as C/Fe to be fairly uniform at any given iron abundance? 
Chemical uniformity requires the interstellar 
medium to be extremely well mixed, but that is precisely what
local spectroscopic studies find. Reddy et al. (2003)
examined this issue based on high-dispersion spectra of a few hundred
stars and  were unable to detect any cosmic scatter. 
The dispersion was as small as 0.03-0.04 dex for many elements, and could be
entirely accounted for considering the 
uncertainties in the atmospheric parameters. 
The immediate implication
is that the local interstellar medium is well mixed and has been well
mixed for many Ga. Such a conclusion is not contradicted by studies
of interstellar gas towards bright stars within and beyond the
local bubble (e.g. Oliveira et al. 2005). 

In this situation it seems only 
natural to expect the Sun to show similar abundance ratios as other
low-mass dwarfs in the solar vicinity with similar metallicity.
That is indeed the case for most elements, but there are some 
striking offsets.
The landmark study by Edvardsson et al. (1993)
found  nearby FGK-type  stars with solar
iron abundance to be, on average, richer than the Sun
in Na, Al, and Si. Part of this trend, but not all, could
be linked to biases in other stellar parameters, such as mean
galactocentric distance and age. 
More recent studies of nearby low-mass stars kept finding 
offsets between the abundance ratios of stars with solar
iron abundances and the Sun. For example, Reddy et al. (2003)
found small offsets, in the same sense as Edvardsson et al.
for the ratios C/Fe, N/Fe, K/Fe, S/Fe, Al/Fe, and Si/Fe 
(and perhaps Na/Fe), but opposite trends 
for Mn/Fe and V/Fe. Allende Prieto et al. (2004) also found 
similar patterns in their  sample for O/Fe, Si/Fe, Ca/Fe, 
Sc/Fe, Ti/Fe,  Ni/Fe, and some neutron-capture elements (Na was not
studied). 

The lack of consistency among different studies regarding the
existence and size of these chemical offsets is worrisome. Local samples
of stars span variable ranges in spectral type, which may be
associated with different systematic errors.
In order to explore further the nature of the observed offsets 
we have observed a sample of solar analogs selected from the
{\it Hipparcos} color-magnitude diagram. We described the results
below.

\begin{figure}
\includegraphics[height=5.5in,angle=90]{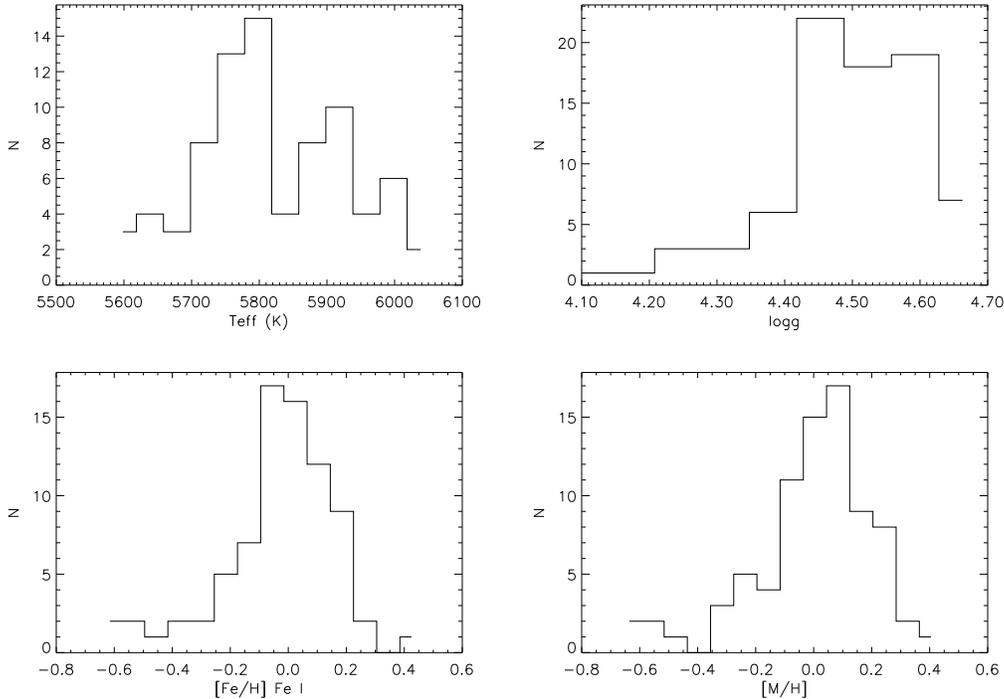}
\caption{
Stellar parameters for the sample. Two {\it metallicity} distributions
are shown: [M/H] indicates the values  derived from the analysis of
the spectral order that includes H$\beta$ (used to select the model
atmosphere), and [Fe/H] indicates the values subsequently derived from
the analysis of equivalent widths of Fe I lines. The surface gravities
shown here correspond to the spectroscopic values derived from the
H$\beta$ order, but the true gravities are likely tightly concentrated
around the solar value ($\log g \simeq 4.437$), given the narrow 
distribution of the sample stars in $M_V$.
\label{f1}
}
\end{figure}

\section{Data and analysis}

To select solar analogs we used the Johnson  $M_V$ absolute magnitudes 
and the $(B-V)$ color indices compiled by Allende Prieto
\& Lambert (1999) for 17,219 nearby stars ($d<100$pc) 
included in the {\it Hipparcos} catalog. Stars were selected
to be within 0.07 mag from the adopted values for the Sun  
$(B-V$,$M_V) = (0.65,4.85)$  and to be accessible to the 9.2m 
Hobby-Eberly Telescope (HET) at McDonald 
Observatory during the first observing period of 2005
(December 2004--March 2005), when the observations were obtained.
A list of 130 stars  were 
placed on the HET queue, and 94 were spectroscopically observed.
 
The observations employed the High Resolution Spectrograph (HRS, Tull 1998),
a fiber-coupled spectrograph, using the first-order diffraction 
grating g316 as cross disperser to get almost continuous
coverage between 407.6	 and 783.8 nm. 
A fiber with a diameter of 
2 arcsec fed the 0.625 arcsec wide slit of the spectrograph, providing
a FWHM resolving power of $R \sim 120,000$.  The data reduction
was made with an automated pipeline within IRAF, performing 
bias correction, flat-fielding, scattered-light correction,  
extraction, and wavelength calibration based on Th-Ar
hollow-cathode spectra. 

The stellar effective temperatures, surface gravities, and overall
metallicity were derived by a $\chi^2$ fitting of the spectral order
containing H$\beta$ (Allende Prieto 2003). First, the procedure
was applied to the spectra of FG dwarfs included as part of the 
Elodie library at a resolving power of $R\sim 10,000$, then the 
residuals were fit by linear trends. After applying the linear corrections,
the rms scatter between our results and those in the Elodie catalog
are 1.5 \%, 0.16 dex and 0.07 dex for $T_{\rm eff}$, $\log g$, and
[Fe/H], respectively. 

\begin{figure}
\includegraphics[height=3.5in,angle=0]{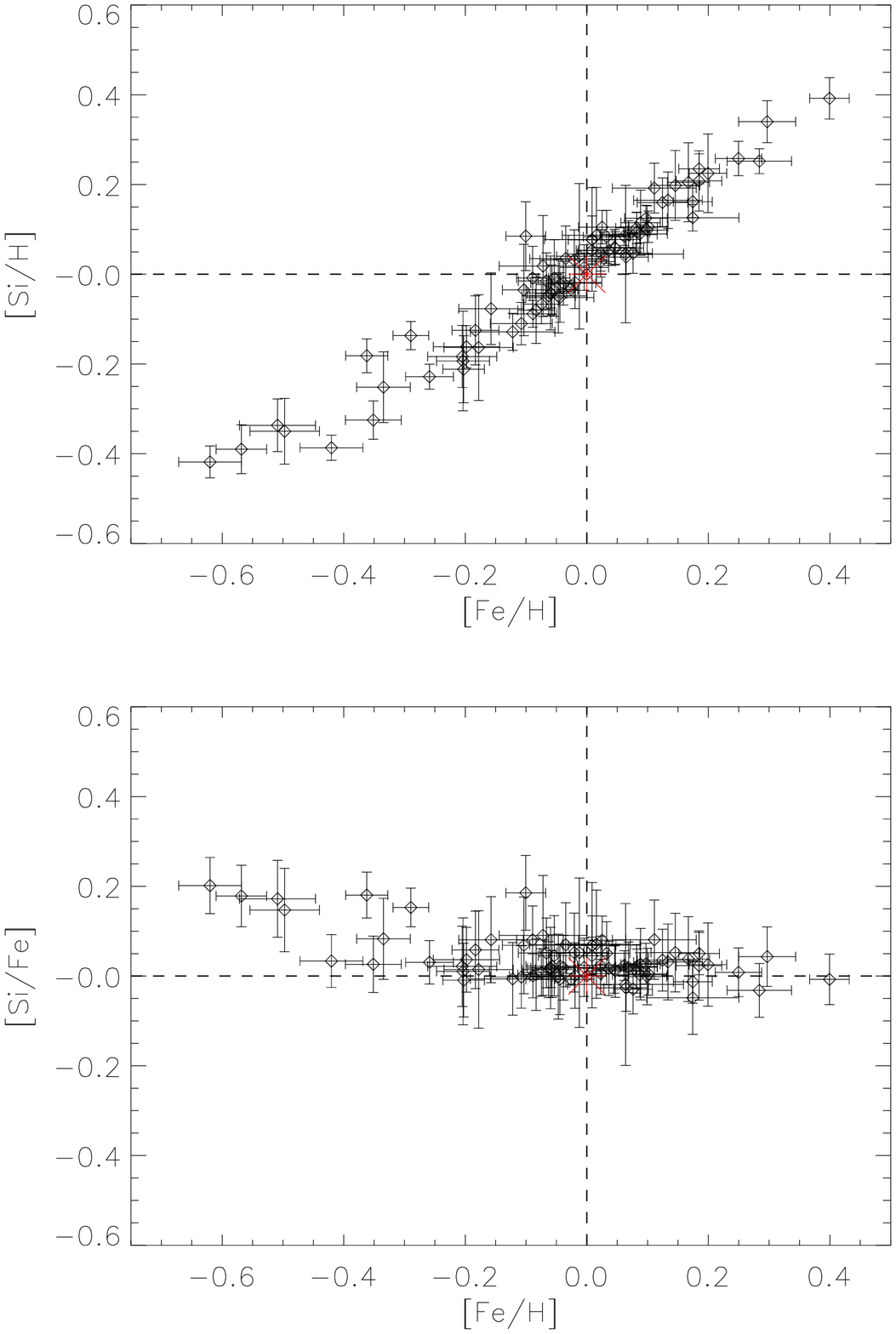}
\hfill
\includegraphics[height=3.5in,angle=0]{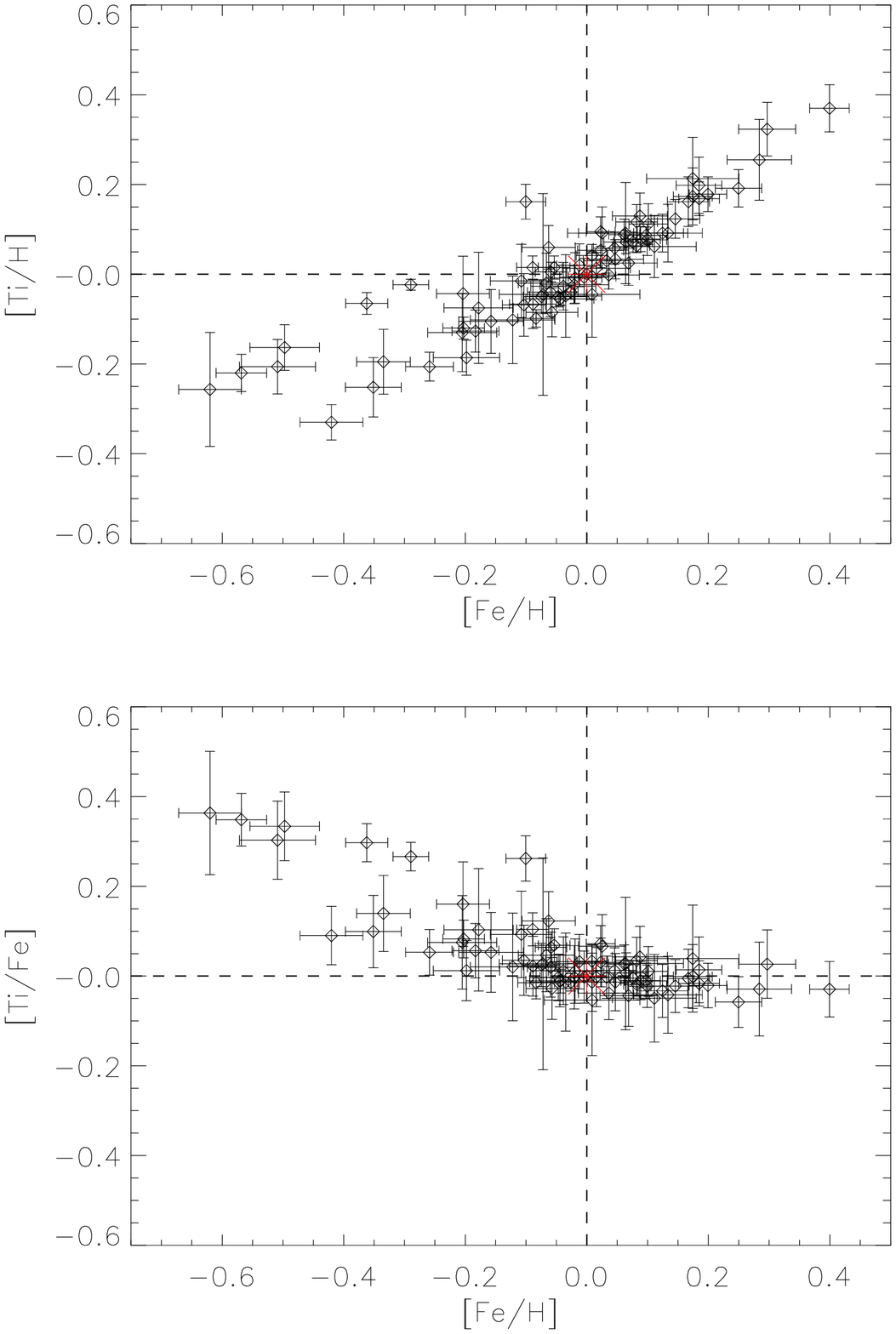}
\caption{
Abundances of Si and Ti, and their ratios to the iron abundances
for the sample stars. The broken lines mark the location of the
solar reference values.
\label{f2}
}
\end{figure}

The HRS spectra were processed in exactly the same manner, after
smoothing them to a resolution $R=10,000$, and the resulting parameters
were subjected to the linear corrections inferred from the comparison with
the Elodie  library.  
Fig. \ref{f1} shows the distribution of the final atmospheric parameters. 
Once the basic atmospheric parameters are constrained, 
we measured and
made use of the equivalent widths of 46 Fe I lines to derive the appropriate
value of the microturbulence and the iron abundance using MOOG 
(Sneden 2002). The iron linelist is
a subset of that described in Ram\'{\i}rez, Allende Prieto \& Lambert (2006),
and for other elements we used the same lines as Allende Prieto et al.
(2004). Abundances of C, Si, Ca, Ti, and Ni were also determined assuming LTE.

\section{Results and discussion}

Fig. \ref{f2} shows our results for silicon and titanium. The thin and
thick disk membership can be easily decided from these plots. The offset
from [Si/Fe]$=0$ at [Fe/H]$=0$ found in several previous surveys is
not apparent in the left-hand panels of Fig. \ref{f2}. Similarly, the
right-hand panels, do not confirm the offset found by Allende Prieto
et al. (2004) for titanium, and no significant 
offsets were found either for carbon, calcium, titanium or nickel.
This result suggests that the offsets in previous analyses were 
likely the result of systematic errors. 

Samples of stars spanning a narrow range in atmospheric parameters
are well-suited to carry out differential studies  of chemical evolution.
With such samples, we can potentially minimize the impact of 
shortcomings in the theory of stellar atmospheres and line formation
on the derived abundances, as the small scatter in the abundance ratios
shown in Fig. \ref{f2} suggest.


\begin{thebibliography}{99}

\bibitem[Allende Prieto(2003)]{2003MNRAS.339.1111A} Allende Prieto, C.\ 
(2003). 
\textit{MNRAS} \textbf{339}, 1111 

\bibitem[Allende Prieto et al.(2004)]{2004A&A...420..183A} Allende Prieto, 
C., Barklem, P.~S., Lambert, D.~L., \& Cunha, K.\ (2004). 
\textit{A\&A} \textbf{420}, 183 

\bibitem[Allende Prieto \& Lambert(1999)]{1999A&A...352..555A} Allende 
Prieto, C., \& Lambert, D.~L. (1999). \textit{A\&A} \textbf{352}, 555 

\bibitem[Bertelli \& Nasi(2001)]{2001AJ....121.1013B} Bertelli, G., \& 
Nasi, E. (2001). 
\textit{AJ} \textbf{121}, 1013 

\bibitem[Edvardsson et al.(1993)]{1993A&A...275..101E} Edvardsson, B.
et al. 
(1993). \textit{A\&A} \textbf{275}, 101 

\bibitem[Haywood(2002)]{2002MNRAS.337..151H} Haywood, M.\ (2002). 
\textit{MNRAS}, \textbf{337}, 151 
         
\bibitem[Nordstr{\"o}m et al.(2004)]{2004A&A...418..989N} Nordstr{\"o}m, 
B., et al. (2004). 
\textit{A\&A} \textbf{418}, 989          

\bibitem[Luck \& Heiter(2005)]{2005AJ....129.1063L} Luck, R.~E., \& Heiter, 
U. (2005).
\textit{AJ} \textbf{129}, 1063 

\bibitem[Oliveira et al.(2005)]{2005ApJ...625..232O} Oliveira, C.~M., 
Dupuis, J., Chayer, P., \& Moos, H.~W.\ (2005). 
\textit{ApJ} \textbf{625}, 232 

\bibitem[]{} Ram\'{\i}rez, Allende Prieto \& Lambert (2007).
\textit{A\&A}, in press

\bibitem[Reddy et al.(2006)]{2006MNRAS.367.1329R} Reddy, B.~E., Lambert, 
D.~L., \& Allende Prieto, C. (2006).
\textit{MNRAS} textbf{367}, 1329 

\bibitem[Reddy et al.(2003)]{2003MNRAS.340..304R} Reddy, B.~E., Tomkin, J., 
Lambert, D.~L., \& Allende Prieto, C. (2003).
\textit{MNRAS} \textbf{340}, 304 

\bibitem[Rocha-Pinto et al.(2000)]{2000A&A...358..869R} Rocha-Pinto, H.~J., 
Scalo, J., Maciel, W.~J., \& Flynn, C. (2000).
\textit{A\&A} \textbf{358}, 869 

\bibitem[]{} Sneden, C. (2002). 
{\tt http://verdi.as.utexas.edu/moog.html}

\bibitem[Vergely et al.(2002)]{2002A&A...390..917V} Vergely, J.-L., 
K{\"o}ppen, J., Egret, D., \& Bienaym{\'e}, O. (2002).
\textit{A\&A}  \textbf{390}, 917 	 

\end{thebibliography}
\end{document}